\documentclass[a4paper,11pt]{article}
\pdfoutput=1 

\usepackage{jcappub} 

\usepackage[T1]{fontenc} 

\usepackage{dcolumn}
\usepackage{braket}
\usepackage{mathtools}
\usepackage{caption}
\usepackage[english]{varioref}
\usepackage{url}

\title{\boldmath Berezinsky Hidden Sources: An Emergent Tension in the High-Energy Neutrino Sky?}

\author[a,b,1]{Antonio Ambrosone,\note{Corresponding author.}}

\affiliation[a]{Gran Sasso Science Institute (GSSI), Viale Francesco Crispi 7, 67100 L’Aquila, Italy}
\affiliation[b]{ INFN-Laboratori Nazionali del Gran Sasso(LNGS), via G. Acitelli 22, 67100 Assergi (AQ), Italy}

\emailAdd{antonio.ambrosone@gssi.it}

\abstract{ The IceCube Collaboration has recently reported compelling evidence of high-energy neutrino emission from NGC~1068,  and also mild excesses for NGC 4151 and CGCG420-015, local Seyfert galaxies. This has increased the interest along neutrino emission from hot-corona surrounding the super massive black holes of Seyfert Galaxies. In this paper, we revisit phenomenological constraints on the neutrino emission from hot-coronae of seyfert galaxies, using an assumption of equi-ripartition between cosmic-rays and magnetic energy densities. We show that not only these sources are consistent with such an assumption but also that the data point towards low beta plasma parameters inside Seyfert Galaxies. We exploit this finding to constrain the Seyfert diffuse neutrino flux and we obtain that, in order not to overproduce neutrinos, not all the sources can be in an equi-ripartition state. We conclude (along with previous findings) that seyfert galaxies cannot explain the diffuse neutrino spectrum above $\sim 100\, \rm TeV$, allowing space for other astrophysical sources.}

\begin{document}
\maketitle
\flushbottom

\section{Introduction}

Seyfert Galaxies are Active Galactic Nuclei (AGNs) powered by super-massive black holes (SMBHs) positioned in their core, whose accretion disk related processes produce very energetic phenomena, such as the creation of a hot corona region surrounding the black hole~\cite{2004MNRAS.351..169M}.  In the late 1970s V.S. Berezinsky proposed that the hot corona region might produce and accelerate high-energy cosmic rays (CRs)~\cite{1981ICRC....1..238B} (see also~\cite{Fiorillo:2023dts} and references therein for further details). these CRs were predicted to predominantly interact with ambient particles leading to the production of $\gamma$-rays and neutrinos, with only the latter being able to escape these very-high dense environments~\cite{1981ICRC....1..238B,Berezinsky:2000bq,Berezinsky:2004kn}. Despite this idea could have not been experimental explored for many years, current $\gamma$-ray and neutrino telescopes have strongly improved our knowledge about astrophysical messengers, making it possible to quantitative constrain this scenario. Firstly, many analyses have highlighted a tension between the diffuse $\gamma$-ray~\cite{Fermi-LAT:2014ryh} and neutrino~\cite{IceCube:2020wum,IceCube:2020acn,IceCube:2021uhz,Abbasi:2024jro} fluxes, especially because of the large neutrino flux below  100 TeV, exploring the potential role of hidden CR accelerators~\cite{Bechtol:2015uqb,Chang:2016ljk,Xiao:2016rvd,Sudoh:2018ana,Capanema:2020rjj,Capanema:2020oet,Murase:2015xka,Fang:2022trf,Murase:2022dog,Inoue:2019fil,Murase:2019vdl} (see also \cite{Ambrosone:2020evo} for other details). Furthermore, the IceCube collaboration has recently found a 4.2~$\sigma$ excess above the background-only hypothesis of 72 high-energy neutrino events coming from the direction of NGC 1068, a nearby AGN~\cite{IceCube:2022der}, 
inferring a power-law differential neutrino flux $(\sim E^{-3.2})$ approximately ten times higher than expected from a $\gamma$-ray transparent source, taking into account the $\gamma$-ray emission observed by the Fermi-LAT telescope from this source~\cite{2012ApJ...755..164A,Ajello:2020zna,Ambrosone:2024xzk}. This, from one hand, has confirmed the idea of the presence of hidden high-energy neutrino sources in the Universe, but from the other hand, it has triggered a lot of controversy for the interpretation of this observation. 
Indeed, a lot of models have been proposed to explain such observations~\cite{Fiorillo:2023dts,Eichmann:2022lxh,Inoue:2019yfs,Murase:2019vdl,Murase:2022dog,Blanco:2023dfp,Kheirandish:2021wkm,Ajello:2023hkh,Padovani:2024tgx,Kun:2024meq,das2024revealing,Yasuda:2024fvc,Fang:2023vdg}. At the moment, the most natural explanation resides in high-energy neutrino emission in the hot corona emission where the high x-ray  photon density is able to strongly suppress the high-energy $\gamma$-ray flux accompanying the neutrino flux, leading to a natural explanation for a $\gamma$-opaque source~\cite{Padovani:2024ibi}.~However, there is not consensus about the acceleration mechanisms for high-energy CRs inside the coronae. For instance, Refs. \cite{Murase:2019vdl,Kheirandish:2021wkm} have proposed stochastic processes as acceleration mechanism while \cite{Inoue:2019fil,Inoue:2019yfs} have proposed diffusive shock acceleration, both normalising the spectrum imposing that the CR pressure is just a small fraction of the thermal gas pressure. On the other hand, Ref.~\cite{Fiorillo:2023dts} recently explored the role of magnetic re-connection, normalising the spectrum considering equi-partition between CR energy density and magnetic energy density. On the contrary, Ref. \cite{Eichmann:2022lxh} ascribed a fraction of the accretion mass rate energy to the production of high-energy CRs. Nonetheless, all the models seemingly share some common properties: firstly, results suggest energy losses be  very important in the hot corona regions. In fact,  Beithe- Heitler pair production leads to electromagnetic cascades shifting the energy of photons in the MeV energy range~\cite{Murase:2019vdl,Murase:2022dog,Murase:2023ccp,Das:2024vug}.
Secondly, the predicted neutrino luminosity of hot corona is expected to be proportional to the AGN X-ray luminosity. Therefore, it is fundamental to observe of other seyfert galaxies to shed light onto the processes occurring in these kind of environments~\cite{Sommani:2024sbp,Neronov:2023aks}. On this regard, IceCube has searched for other seyfert galaxies in the local Universe, both as individual emitters and performing a stacking search \cite{Abbasi:2024ofy,Abbasi:2024hwv,IceCube:2023tts,IceCube:2023nai}. 
The authors found a $\sim 3\sigma$ post-trial excess in the vicinity of NGC 4151~\cite{Abbasi:2024hwv}  with spectral index $\gamma = 2.83_{-0.28}^{+0.35}$.  Furthermore, an excess has been found also in the vicinity of CGCC 420-015~$(2-2.5~\sigma)$~\cite{Abbasi:2024ofy}, with a spectrum very similar to NGC 1068 spectrum.  
 By constrast, Ref.~\cite{Neronov:2023aks}, analysing 10 years of public IceCube data, found an excess even in the direction of NGC 3079, another seyfert galaxy. Seemingly, these potential observations point towards a correlation between  the neutrino and hard x-ray luminosity for these sources, although the results are too premature to draw robust conclusions~\cite{Kun:2024meq}.
In fact, while Refs.~\cite{Murase:2023ccp,Das:2024vug} have produced optimistic estimates for the neutrino emission of several local seyfert galaxies, Ref. \cite{Inoue:2024nap} has, on the contrary,  argued that the seyfert hot coronae should not be able to produce a neutrino flux as high as the one observed by the IceCube collaboration. Furthermore, it is not clear if the steep spectra measured are byproducts of copious p$\gamma$ interactions happening in the hot coronae or if they directly represent the cut-off of the injected CR fluxes \cite{Das:2024vug}. In this paper, we derive new data-driven constraints on the seyfert neutrino emission exploiting all the state-of-the-art observations of the IceCube collaboration also exploring the potential role played by these sources to diffuse neutrino flux.
To such a purpose, we employ a model which takes into account both CR escape mechanisms as well as CR energy losses and consistently solve the CR transport equation inside the hot-coronae. We normalise the spectrum assuming equi-ripartition between CR and magnetic energy densities, which represent an absolute maximum value for the CR normalization. We also ensure that the injected CR luminosity is only a fraction of the bolometric AGN luminosity which is self-consistently calculated through the background photon energy density. We show that the four sources which provide an excess into IceCube data are consistent with an equi-ripartion between CRs and magnetic density. In fact, the fit prefers small beta plasma values pointing towards high magnetic field values and a rather high energetic carried by CRs into Seyfert Galaxies~(see below for further details). We then extrapolate this information to the whole seyfert population (diffuse neutrino flux), using the X-ray luminosity function (see \cite{2014ApJ...786..104U}), showing that it slightly overproduce the diffuse neutrino flux inferred by the ICeCube collaboration through the starting track sample~\cite{Abbasi:2024jro} in the $\sim 1-10\, \rm TeV$ range. Future observations from the upcoming neutrino telescopes, such as KM3NeT~\cite{KM3NeT:2024uhg}, IceCube gen 2~\cite{IceCube-Gen2:2020qha}, P-ONE~\cite{P-ONE:2020ljt} and the TRIDENT~\cite{Ye:2023dch} telescopes will be fundamental to unveil the role of seyfert galaxies to the diffuse neutrino spectrum as well as the role of other astrophysical accelerators. The paper is organised as follows: in Sec. \ref{sec:model}, we describe the CR transport model and  the neutrino emission of seyfert galaxies, in Sec. \ref{sec:observations}, we summarise the X-ray and neutrino observations for the four sources observed by IceCube. In sec.~\ref{sec:statistical_analysis}, we quantitative test our model using the observations and report the results. In Sec.~\ref{sec:diffuse}, we discuss the role of seyfert galaxies into the diffuse neutrino spectrum and in Sec.~\ref{sec:conclusions} we draw our conclusions. Finally, in appendix~\ref{sec:timescales}, we provide details on the dynamical timescales in the hot-coronae and the neutrino production efficiency for proton-proton collisions and photomeson interactions.

\section{On the Neutrino Emission of Seyfert Galaxies}\label{sec:model}

The hot coronae regions are generally very small and extend only for 1-100 Schwarzschild radii~$(\mathcal{R}_S)$ around the SMBHs~\cite{Murase:2023ccp}. In this work, we use spherical geometry with a radius $R = r \cdot \mathcal{R}_S$, fixing $r=20$ (consistent with the recent upper limits obtained by Ref.~\cite{Das:2024vug} with sub-GeV gamma-rays data). The CR transport equation reads
\begin{equation}\label{eq:transport}
    \frac{N_{CR}(E)}{\tau_{\rm esc}} - \frac{d}{dE} \bigg[\frac{E}{\tau_{\rm loss}} N_{CR}(E)\bigg] = Q(E)
\end{equation}
where $\tau_{\rm esc}$ is the escape timescale, $\tau_{\rm losses}$ is the energy loss timescale (see app.~\ref{sec:timescales} for details about the modelling on timescales) and finally $Q(E)$ is the injection rate of CRs. Eq.~\ref{eq:transport} is solved through the green function as (see for instance \cite{Inoue:2019fil})
\begin{equation}
    N_{CR}(E) = \frac{\tau_{\rm loss}}{E}\int_{E}^{+\infty} Q(E_1) e^{-G(E,E_1)} dE_1
\end{equation}
with 
\begin{equation}
    G(E,E_1) = \int_{E}^{E_1} dE_2  \frac{\tau_{\rm loss}(E_2)}{E_2 \cdot \tau_{\rm esc}} 
\end{equation}

The injection CR rate is assumed to be a power-law with an exponential cut-off at $E_{\rm max} = 200\, \rm TeV$, namely $Q(E) = A (\frac{E}{m_p c^2})^{-\gamma} e^{-E/E_{\rm max}}$. The normalisation parameter $A$ is fixed so that the CR energy density $U_{CR}$ is in equi-ripartion with the magnetic energy density $U_B$, namely $U_{CR} = U_B$, with
\begin{equation}
    U_{CR} = \int_{m_p c^2}^{+\infty} E N_{CR} (E) dE 
\end{equation}
and $U_B = B^2/8\pi$. The magnetic field is set as \cite{Murase:2019vdl}
\begin{equation}
    B = \sqrt{\frac{8\pi n_p K_B T_p}{\beta}}
\end{equation}
where $K_B$ is the Boltzmann constant, $T_p$ is the thermal proton virial temperature and finally $\beta$ is the plasma parameter which represents the ratio between the thermal and magnetic pressure \cite{Murase:2019vdl}. The proton virial temperature is  $m_p c^2/(K_B\cdot r) \simeq 6.0\cdot 10^{10} (r/30)^{-1} \, [\rm K] $ \cite{Murase:2019vdl}. It is important to take into account that the luminosity injected in CR is just a fraction of the bolometric luminosity of the AGN disk~$(L_{\rm ph})$. Therefore, we also constrain the normalization parameter $A$ with
\begin{equation}\label{eq:norm_max}
    \int_{m_p c^2}^{+\infty} E Q(E) dE \le \eta L_{\rm ph}
\end{equation}
 The bolometric luminosity gets contribution both from  the accretion disk in the ultraviolet (UV) regime (usually also called the blue bump of AGNs) and from the coronae in the x-ray luminosity~\cite{Inoue:2019fil,Murase:2019vdl,Fiorillo:2023dts}. In this work, we utilise the model put forward by Ref.~\cite{Inoue:2019fil}, where the number density of background photons can be modelled  in terms of the x-ray luminosity of the AGN in the $[2-10]\, \rm KeV$ band (here forth denominated simply $L_X$) (see \cite{Inoue:2019fil} for more details). We fix $\eta = 25\%$ for the maximal efficiency conferred into CRs.  In order to calculate the pp neutrino production rate, we employ the analytical prescription of ref.~\cite{Kelner:2006tc}
\begin{equation}
    Q_{\nu+\bar{\nu}}^{pp}(E) =\frac{1}{3} c\cdot n_{p}\int_{10^{-3}}^{1} \sigma_{pp}\big(\frac{E}{x}\big) N_{CR}\big(\frac{E}{x}\big) \Tilde{F}_{\nu}\big(x,\frac{E}{x}\big) \frac{dx}{x}
\end{equation}
where $ \Tilde{F}_{\nu}(x,\frac{E}{x})$ is defined in Ref.~\cite{Kelner:2006tc} and it takes into account all the neutrinos produced in the interaction, while the factor 1/3 takes into account neutrino oscillation. Since for $E_{CR}\lesssim 200\, \rm TeV$, the neutrino production is dominated by pp collisions (see appendix~\ref{sec:timescales} for details), for photomeson neutrino production, we estimate the neutrino production rate with multi-messenger relations~\cite{Murase:2022dog,Pisanti:2019zji}
\begin{equation}\label{eq:neutrinos}
    E_{\nu}^2 Q_{\nu + \bar{\nu}}^{p\gamma} \simeq \frac{1}{8} \bigg[\frac{E_{\rm CR}^2 N_{\rm CR}(E)}{\tau_{p\gamma}}\bigg]|_{E_{\rm CR} \simeq 20E_{\nu}}
\end{equation}

we evaluate the neutrino luminosity of the sources between $1-10\, \rm TeV$ as
\begin{equation}\label{eq:Lnu_model}
    L_{\nu +\bar{\nu}}^{1-10\, \rm TeV} = \int_{1\, \rm TeV}^{+\infty} E_{\nu} Q_{\nu + \bar{\nu}}^{\rm tot}(E_{\nu}) dE_{\nu}
\end{equation}
with $Q_{\nu + \bar{\nu}}^{\rm tot}(E)=Q_{\nu + \bar{\nu}}^{pp}(E) +Q_{\nu + \bar{\nu}}^{p\gamma}(E)$. The differential neutrino flux at Earth is 
\begin{equation}
    \phi_{\nu + \bar{\nu}}(E) = \frac{(1+z)^2}{4\pi D_l(z)^2} Q_{\nu + \bar{\nu}}^{\rm tot}(E(1+z))
\end{equation}
where $z$ and $D_l(z)$ are respectively the redshift and the luminosity density.

\section{Neutrino vs X-ray Luminosity}\label{sec:observations}

In this section, we summarise the observations regarding the four neutrino sources which give an excess into IceCube data. For consistency and in order to avoid any bias, we use the values reported \cite{2017ApJS..233...17R} in the BASS AGN catalogue for all the sources for redshifts and the x-ray luminosities. In particular, regarding the x-ray flux, we use the intrinsic x-ray (absorption-corrected) in the 2-10 KeV band. Tab. \ref{tab:values_sources_BASS} summarises all the values considered.

\begin{table}[h!]
    \centering
    \begin{tabular}{c|c|c|c|}
        Source & redshift (z) &  $D_l (z) (\rm Mpc)$ & $F_X (10^{-12}\, \rm erg \, \rm cm^{-2}\, \rm s^{-1})$   \\
        NGC 1068 &  $3.03\cdot 10^{-3}$& 13.4 & $268.30$ \\
        NGC 4151 & $3.14\cdot 10^{-3}$ & 13.9 & 84.80 \\
        CGCG 420-015 & $2.959\cdot 10^{-2}$ & 133.9 & 50.50 \\
         NGC 3079 & $3.40\cdot 10^{-3}$ & 15.1 & 6.60 \\
    \end{tabular}
    \caption{The table summarises the value we use for redshift, luminosity distance and the intrinsic x-ray fluxes of the sources. }
    \label{tab:values_sources_BASS}
\end{table}
we compute the luminosity distance through redshift and using the standard cosmological parameters $\Omega_{M} = 0.31$, $\Omega_{\Lambda}$ and $H_0 = 67.74\, \rm Km\, \rm s^{-1}\, \rm Mpc^{-1}$  \cite{Planck:2018vyg}. We compute $L_X$ as~\cite{2017ApJS..233...17R}
\begin{equation}\label{eq:xrayluminosity}
    L_X = F_x 4\pi D_l^2(z) (1+z)^{2-\Gamma}\simeq F_x 4\pi D_l^2(z)
\end{equation}
where we can neglect the contribution coming from redshift because these sources are very near and their x-ray spectrum is near to a $E^{-2}$ dependence of the x-ray flux. 

It is paramount to correctly assess the uncertainty affecting the values reported in Tab. \ref{tab:values_sources_BASS}. Firstly, there is uncertainty in the distance value;  for instance, Refs. \cite{Padovani:2024tgx,Fiorillo:2023dts} report $10.1\, \rm Mpc$ as the best quoted value for the distance of NGC~1068. We conservatively consider a 0.13 dex uncertainty in the distance in order to be consistent with this estimate. Furthermore, there is an uncertainty on the intrinsic x-ray flux, since the geometry of the absorption region in AGNs is unknown~\cite{Marinucci:2015fqo}. In fact, estimates on $L_X$ for NGC 1068 span from $\sim 5\cdot 10^{42}\, \rm erg\, \rm s^{-1}$ \cite{2017ApJS..233...17R} to $\sim 5\cdot 10^{43}\, \rm erg\, \rm s^{-1}$~\cite{Marinucci:2015fqo}. Therefore,  we consider an uncertainty of 0.7 dex for $F_X$ in order to obtain an overall range for $L_X$ consistent with both estimates (we sum the uncertainty quadratically) leading to a total 0.75 dex uncertainty on $\log L_X$.

\begin{figure}[h!]
    \centering
    \includegraphics[width=\columnwidth]{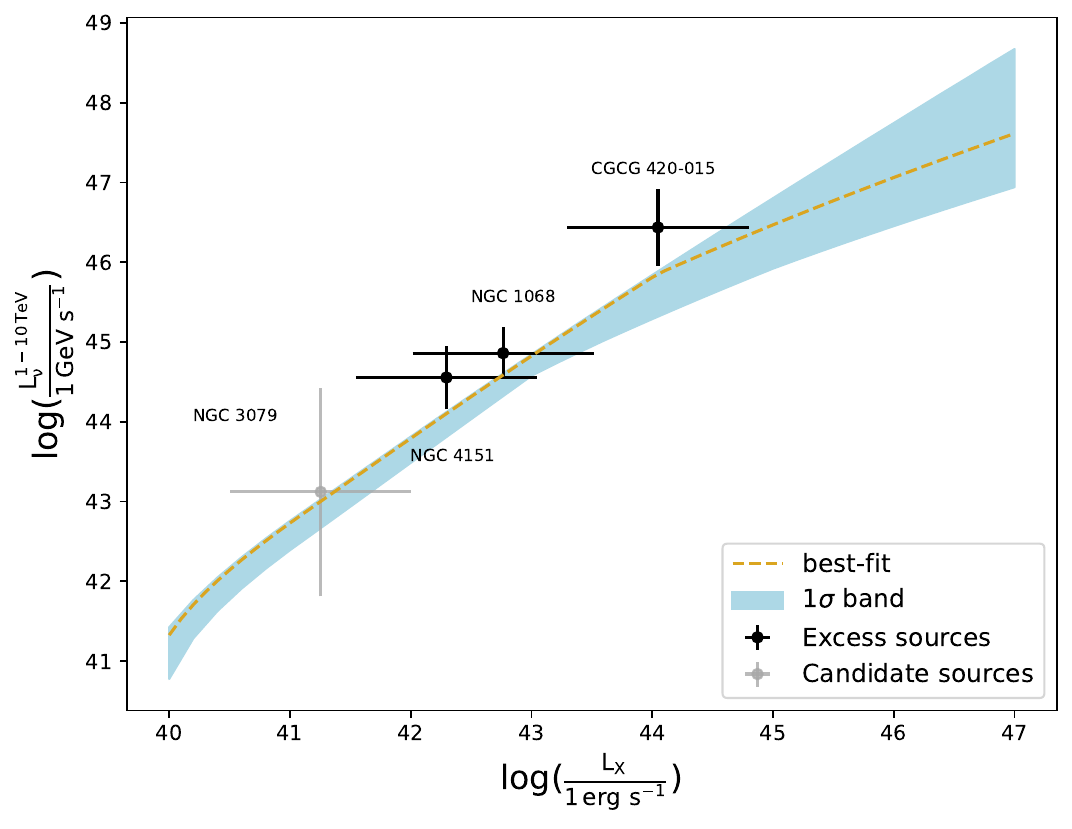}
    \caption{$L_{\nu}^{(1-10\, \rm TeV)}$ as a function of the $L_X$. The uncertainty on $L_{\nu}$ comprises distance and statistical uncertainty, while the uncertainty on $L_X$ is set at 0.75 dex in order to be consistent with all the $L_X$ values inferred in the literature. NGC~1068, NGC~4151, CGCG~420-015 are shown as black datapoints while NGC~3079 as a grey datapoint.
     The blue band corresponds to the $1\sigma$ uncertainty band from our statistical analysis, while the golden dashed line corresponds to the best-fit scenario.  }
    \label{fig:Lnu_vs_Lx_model}
\end{figure}

For the neutrinos, we consider the energy range $1-10\, \rm TeV$ which is interval where IceCube is most sensitive to very steep spectra observed for these sources~\cite{IceCube:2022der}. Therefore, we compute the luminosity as
\begin{equation}
    L_{1-10\, \rm TeV} = 4\pi D_l(z)^2 (1+z)^{2-\gamma} \int_{1\, \rm TeV}^{10\, \rm TeV} E \phi_{\nu}(E,\gamma) dE \simeq  4\pi D_l(z)^2 \int_{1\, \rm TeV}^{10\, \rm TeV} E \phi_{\nu}(E,\gamma) dE 
\end{equation}
where we neglect the dependence over z, because all the sources reside in the very local Universe. For NGC 1068, NGC 4151 and CGCG420-015, we consider the best-fit and the 68\% CL contours for the power-law fit recently published by the IceCube collaboration~\cite{Abbasi:2024hwv,Abbasi:2024ofy} (see also \cite{IceCube:2022der,IceCube:2023tts,IceCube:2023nai}); while for   For NGC 3079, we utilise the spectral energy distribution~(SED) shown by~\cite{Neronov:2023aks}. For the uncertainty, we consider the statistical $68.3\%$ CL uncertainty as well as the distance uncertainty. Fig. \ref{fig:Lnu_vs_Lx_model} shows $L_{\nu}^{1-10\, \rm TeV}$ in terms of $L_X$ for all the sources.  We report NGC 1068, NGC 4151 and CGCG420-015 as black datapoints, while NGC 3079 is reported as grey data point.

\section{Statistical Analysis and Results}\label{sec:statistical_analysis}

In this section, we test if our theoretical model~(\ref{sec:model}) can follow the experimental results outlined in the previous section~\ref{sec:observations}.
For this purpose, we define the following chi-square
\begin{equation}
    \chi^2(\gamma, \beta) = \sum_{i} \frac{(\ log L_{\nu}^{\rm obs, i} - \rm log L_{\nu}^{\rm mod} (\rm log L_X^{\rm obs, i} \gamma, \beta))^2}{\sigma_{\ log L_{\nu}}^2+ \bigg(\frac{\partial \rm log L_{\nu}^{\rm mod} (\rm log (L_X^{\rm obs, i}), \gamma, \beta)) }{\partial  \ log L_X}\bigg)^2 \sigma_{\ log_{L_X}}^2 } 
\end{equation}
The $\partial \ log L_{\nu} / \partial L_X$ term allows us to take into account also the uncertainty on $L_X$. We use $\Delta \chi^2 = \chi^2(\gamma, \beta) - \rm min \, \chi^2 $ as a mean to probe the parameter space for $\gamma \in [0,3]$ and $\beta \in [10^{-2},10]$. Consistently with the Wilks's theorem~\cite{Wilks:1938dza}, the $\Delta \chi^2$ distribution is  a chi-squared with a degree of freedom equal to the number of free parameters in the fit~(2).

\begin{figure}[h!]
    \centering
    \includegraphics[width=\columnwidth]{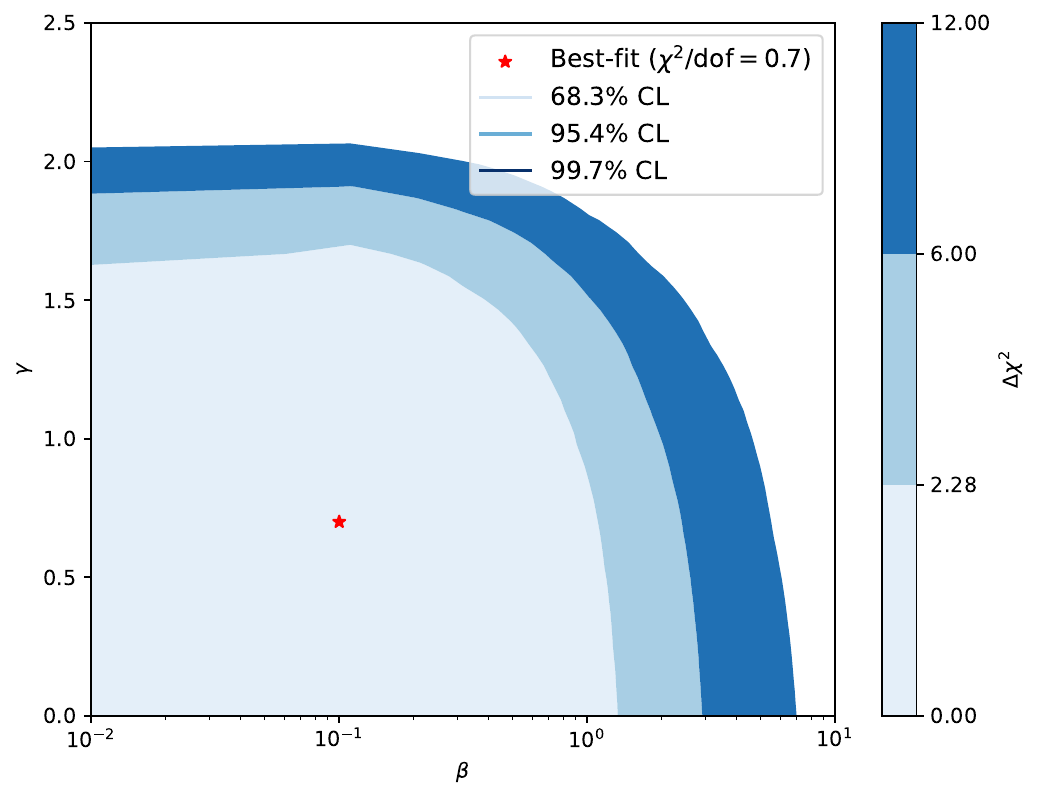}
    \caption{$\Delta \chi^2$ contours for our fit. the color scheme represents, respectively, the contours at $1,2$ and $3$ $\sigma$. The red star represents the best fit scenario.}
    \label{fig:contours_model}
\end{figure}

\begin{figure}
    \centering
    \includegraphics[width=0.49\columnwidth]{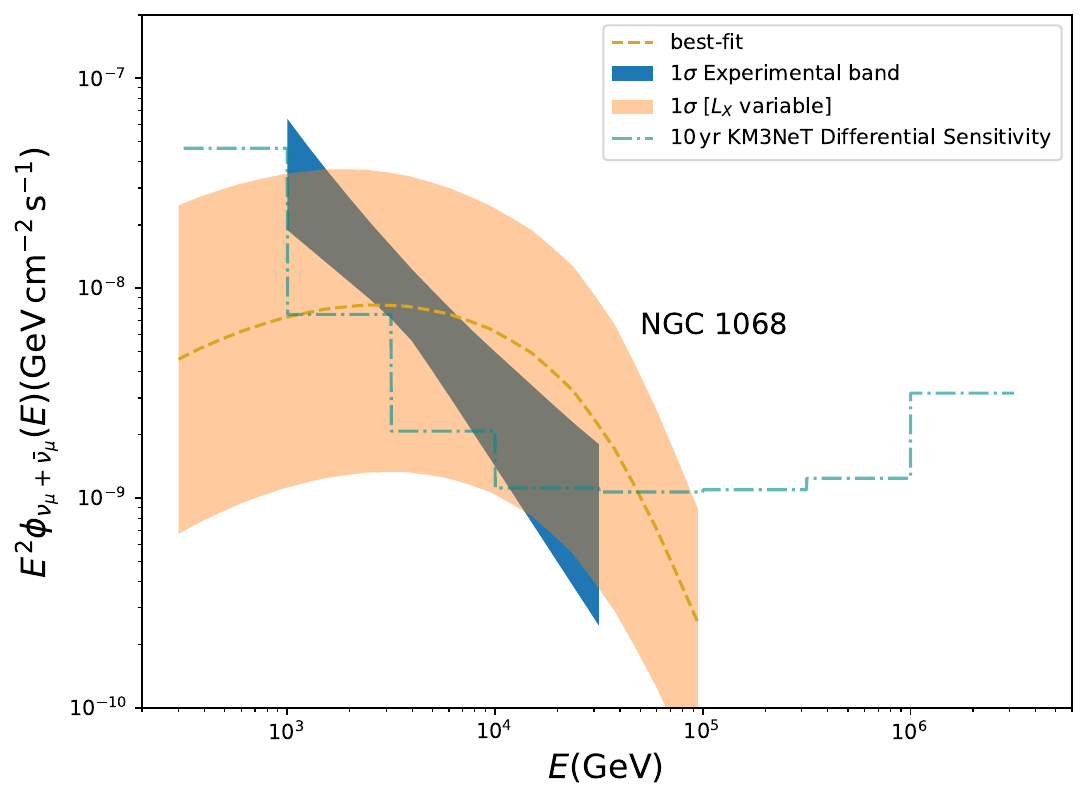}
      \includegraphics[width=0.49\columnwidth]{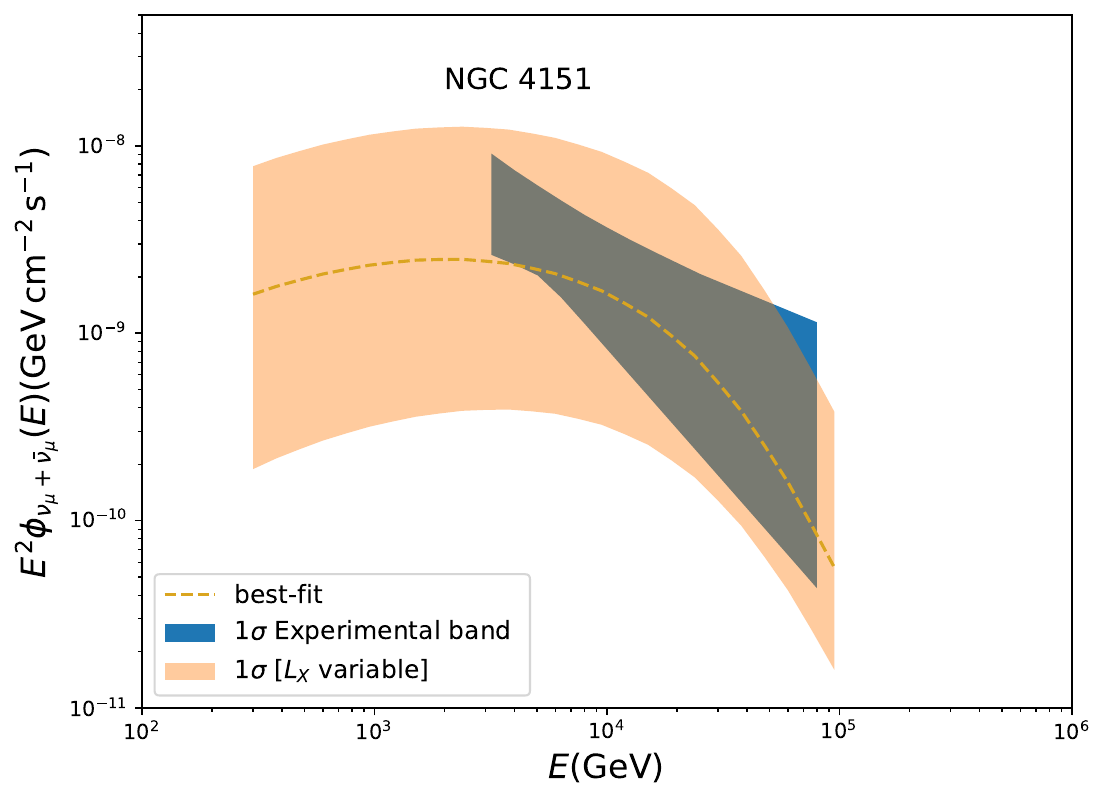}
    \caption{\textbf{Left}:  $1\sigma$ SED band measured by IceCube~\cite{IceCube:2022der,Abbasi:2024hwv}~(blue band) for NGC 1068 compared with predictions of our statistical analysis. In particular, we show the best-fit scenario (golden dashed line) and the  $1\sigma$ band allowing $L_X$ to vary within the uncertainty shown in Fig.~\ref{fig:Lnu_vs_Lx_model} and fixing $D_L$ to the values shown in Tab.~\ref{tab:values_sources_BASS} (orange band). We also report the expected KM3NeT differential sensitivity after 10 years of full operation~\cite{KM3NeT:2024uhg}. \textbf{Right}: the same as on the left but for NGC 4151. The SED is directly taken from \cite{Abbasi:2024hwv} (see also \cite{IceCube:2023tts,IceCube:2023nai}). }
    \label{fig:SED_sources_1}
\end{figure}

\begin{figure}
    \centering
            \includegraphics[width=0.48\columnwidth]{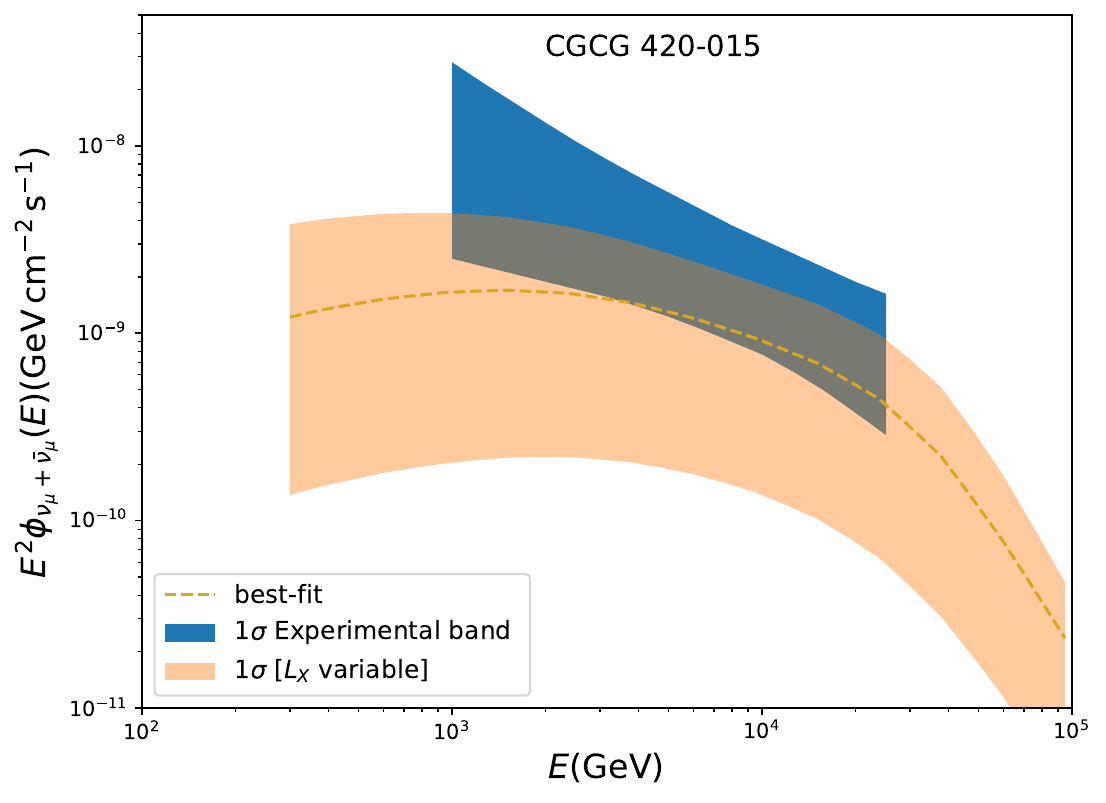}
        \includegraphics[width=0.48\columnwidth]{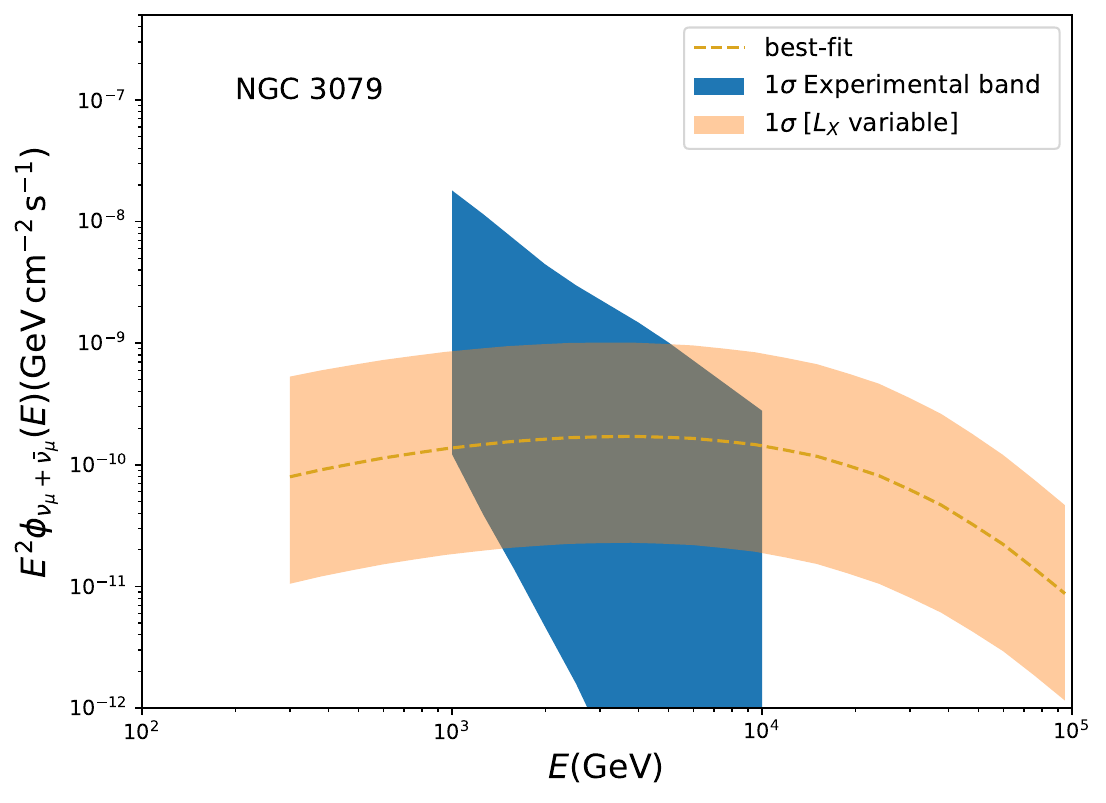}
    \caption{\textbf{Left}:  $1\sigma$ SED band for CGCG420-015  (blue band)  \cite{Abbasi:2024ofy,IceCube:2023tts,IceCube:2023nai})) compared with predictions of our statistical analysis. In particular, we show the best-fit scenario (golden dashed line) and the  $1\sigma$ band allowing $L_X$ to vary within the uncertainty shown in Fig.~\ref{fig:Lnu_vs_Lx_model} and fixing $D_L$ to the values shown in Tab.~\ref{tab:values_sources_BASS} (orange band). \textbf{Right}: the same as on left but for NGC 3079. The SED is taken from~\cite{Neronov:2023aks}.}
    \label{fig:SED_sources_2}
\end{figure}

Therefore, the exclusion limits at 1,2 and $3\sigma$ are respectively defined as  $\Delta \chi^2 = 2.28, 6, 12$. Fig.~\ref{fig:contours_model} shows the best-fit scenario as well as the  1,2 and $3\sigma$ contours. In Fig.~\ref{fig:Lnu_vs_Lx_model}, we also report the $1\sigma$ band for $L_{\nu}$ as well as the best-fit scenario obtained.  We also compare our predicted SEDs with the ones obtained by IceCube~(see Figs. \ref{fig:SED_sources_1} and \ref{fig:SED_sources_2}). In particular, we show the best-fit scenario (golden dashed line) and the $1\sigma$ band (orange) allowing for $L_X$ to vary within the uncertainty reported in Fig.~\ref{fig:Lnu_vs_Lx_model} fixing $D_L$ to the values shown in Tab.~\ref{tab:values_sources_BASS}. For NGC 1068, we also report the expected KM3NeT differential sensitivity after 10 years of full operation~\cite{KM3NeT:2024uhg}. 

 Despite all the uncertainties,  the results allow us to derive some general conclusions: 
the four sources are consistent with the assumption of equi-ripartition between CRs and the magnetic energy density. In fact, low values for $\beta~(\lesssim 1)$~(\cite{Das:2024vug})~are preferred by the data in order to increase the magnetic field and so the CR flux normalisation. However, Ref.~\cite{Inoue:2024nap} has argued that in seyfert galaxies might be nonphysical to get low beta parameters since they are not usually characterised by strong jets. Along with previous findings~\cite{Fiorillo:2023dts,Murase:2022dog,Murase:2023ccp,Das:2024vug}, very hard spectral indexes are needed in order to explain the measurements. Indeed, $\gamma \simeq 0.7$ as a best-fit value is very near to the value predicted by stochastic CR acceleration mechanism~\cite{Kheirandish:2021wkm,Murase:2022dog} and magnetic reconnection~\cite{Fiorillo:2023dts}. In general, $\gamma \gtrsim 1.9 $ are in tension at $\sim 2\sigma$ with current observations, because softer injected spectra would need a higher normalisation to fit the observations. On this regard, we stress that with lower $\eta$ values in Eq.~\ref{eq:norm_max}, the model fails to completely fit all the data especially for NGC 1068 and CGCG420-015, which means that if the neutrino production indeed takes place in the hot corona, an enormous energetics is carried by CRs. Recently, Ref.~\cite{Das:2024vug} has proposed that CRs are only accelerated in specific energy ranges $(\sim 10-200\, \rm TeV)$ in order to reduce the tension with energetics. We leave this scenario for future exploration. Finally, we stress  that for CGCG420-015  the model can barely be consistent with the observations. 

\section{Diffuse Spectrum}\label{sec:diffuse}

Equipped with the previous results, in this section,  we extrapolate the information to the whole population constraining the diffuse neutrino fluxes of seyfert galaxies. 

For the distribution, we follow the distribution of Ref.~\cite{2014ApJ...786..104U} (see \cite{Inoue:2019fil} for further details)
and define the comoving density of sources as 
\begin{equation}
   \rho(L_X,z) = \frac{d \Phi_X(L_X,z)}{d \ log L_X} = \frac{d \Phi_X (L_X,0)}{d \ log L_X} e(z,L_X)
\end{equation}

with 
\begin{equation}
    \frac{d \Phi_X(L_X,0)}{d \ log L_X} = A \bigg[\bigg(\frac{L_X}{L_{*}}\bigg)^{\gamma_1} + \bigg(\frac{L_X}{L_{*}}\bigg)^{\gamma_2}\bigg]^{-1}
\end{equation}
and 
 \begin{equation}
    e(z,L_x) =
    \begin{cases*}
      (1+z)^{p_1} & if $z \le z_{c_1}(L_X) $ \\
      (1+z_{c_1})^{p_1} \big(\frac{1+z}{1+z_{c_1}}\big)^{p_2}        & if $z_{c_1}(L_X) \le z\le z_{c_2}(L_X)$ \\
   (1+z_{c_1})^{p_1} \big(\frac{1+z_{c_2}}{1+z_{c_1}}\big)^{p_2} \big(\frac{1+z}{1+z_{c_2}}\big)^{p_3}    &  if $z\ge z_{c_2} $
       \end{cases*}
\end{equation}

where 
\begin{equation}
    p_1 (L_X) = p^{*}_{1} + \beta_1 (\ log L_X -44)
\end{equation}
while 

\begin{equation}
 z_{c_1} (L_X) = 
    \begin{cases*}
        z_{c_1}^{*} \big(\frac{L_X}{L_{a_1}}\big)^{\alpha_1} & if $L_X \le L_{a_1}$ \\
        z_{c_1}^{*} & otherwise
    \end{cases*}
\end{equation}

and 

\begin{equation}
 z_{c_2} (L_X) = 
    \begin{cases*}
        z_{c_2}^{*} \big(\frac{L_X}{L_{a_2}}\big)^{\alpha_2} & if $L_X \le L_{a_2}$ \\
        z_{c_2}^{*} & otherwise
    \end{cases*}
\end{equation}

Tabs.~\ref{tab:dist_parameters_1} and \ref{tab:dist_parameters_2} summarise  all the parameters used in the distribution (see \cite{2014ApJ...786..104U} for further details).

\begin{table}[h!]
    \centering
    
       \begin{tabular}{c|c|c|c|c|c|c|c|c|c|}
        $A \small{(10^{-6} \, h_{70}^3\, \rm Mpc^{-3})}$ & $\log L_{*}$ & $\gamma_1$ & $\gamma_2$ & $p_1^{*}$ & $\beta_1$ \\
       $2.91 \pm 0.07$  & $43.97 \pm 0.06$ & $0.96 \pm 0.04$ & $2.71 \pm 0.09 $ & $4.78 \pm 0.16$ & $0.84 \pm 0.18$ 
    \end{tabular}
    \caption{Summary of the distribution parameters taken from \cite{2014ApJ...786..104U}. All the luminosity are expressed in units of $\rm erg\, \rm s^{-1}$.}
    \label{tab:dist_parameters_1}
\end{table}

\begin{table}[h!]
    \centering
       \begin{tabular}{c|c|c|c|c|c|c|c|c|c|}
        $z_{c_1}^{*}$ & $\log L_{a_1}$& $\alpha_1$ & $p_2$ & $p_3$ & $z_{c_2}^{*}$ & $\log L_{a_2}$ & $\alpha_2$  \\
      $1.86 \pm 0.07$ & $44.61 \pm 0.07$ & $0.29\ \pm 0.02$ & -1.5 & -6.2 & 3.0 & 44 & -0.1
    \end{tabular}
    \caption{Continuation of Tab.~\ref{tab:dist_parameters_1} for the distribution parameters.  All the luminosity are expressed in units of $\rm erg\, \rm s^{-1}$.}
    \label{tab:dist_parameters_2}
\end{table}

\begin{figure}[h!]
    \centering
    \includegraphics[width=\columnwidth]{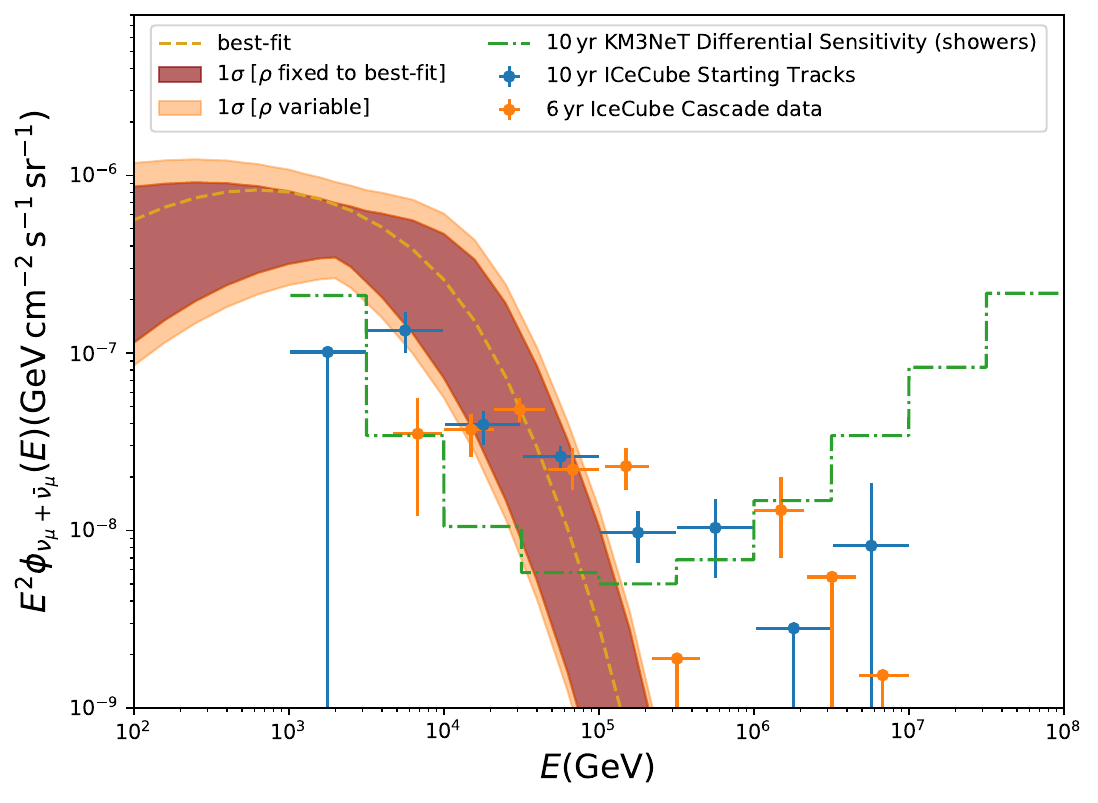}
    \caption{best-fit diffuse neutrino flux from seyfert galaxies (golden dashed line), the $1\sigma$ band fixing the source density distribution to the best-fit scenario (dark red scenario) and the total $1\sigma$ band considering an uncertainty over the density distribution (orange band) compared compared with the diffuse IceCube fluxes (10 years of starting tracks \cite{Abbasi:2024jro} and 6 years of cascades flux \cite{IceCube:2020acn}) and with the expected KM3NeT differential sensitivity for shower events~\cite{KM3NeT:2024uhg}. }
    \label{fig:diffuse}
\end{figure}

Along with Ref.~\cite{Inoue:2019fil}, we multiply the A parameter for 1.5 in order to take into account the fraction of compton-thick AGNs into the x-ray luminosity function. 
The final diffuse neutrino spectrum per solid angle reads 
\begin{equation}\label{eq:final_flux_diffuse}
    \Phi_{\nu}(E_{\nu}, \gamma, \eta, E_{\max}) = \frac{c}{4\pi H_0} \int_{0}^{z_{\rm max}} \frac{dz}{E(z)} \int_{10^{41}}^{10^{47}}    \rho(L_X,z) Q_{\nu + \bar{\nu}}(E_{\nu}(1+z),L_X,\gamma,\eta,E_{\rm max}) 
\end{equation}

with $E(z) = \sqrt{\Omega_M (1+z)^3+ \Omega_\Lambda}$ and the $L_X$ integration limits are set to be $10^{41}\, \rm erg \, \rm s^{-1}$ and $10^{47}\, \rm erg \, \rm s^{-1}$. Fig.~\ref{fig:diffuse} shows our final best-fit scenario (golden dashed line), the $1\sigma$ band according to our stastical analysis fixing the source distribution density to the best-fit scenario (dark red band) and finally the maximal $1\sigma$ band taking also into account the uncertainty on the density distribution (orange band). We compare the results with the latest diffuse IceCube data (6 year cascade \cite{IceCube:2020acn} and 10 year of starting tracks \cite{Abbasi:2024jro}) and the expected KM3NeT differential sensitivity after 10 years of full operation considering all-sky shower events~\cite{KM3NeT:2024uhg}. 

We notice that if we assume that all the seyfert galaxies have the same relation between $L_{\nu}$ and $L_X$ as imposed by the four sources used in the statistical analysis, we  overproduce the diffuse neutrino flux at $\sim 1-10\, \rm TeV$. On this regard, the latest diffuse flux from  IceCube starting tracks strongly constrain the astrophysical diffuse neutrino spectrum at $\sim 1 \, \rm TeV$~\cite{Abbasi:2024jro}. We stress that this result is independent on the possible energetics tension given by the low $\beta$ plasma parameter because $L_{\nu}$ and $L_X$ are data-driven and any model should be able to reproduce these, leading to almost the same result. This has already been emphasised by Ref.~\cite{Padovani:2024tgx} where the authors has evaluated the cosmic neutrino background from non-jetted AGNs, assuming that all AGNs behave like NGC 1068. 
Our best-fit result is slightly higher than the one reported in Ref.~\cite{Padovani:2024tgx} because our model self-consistently exploits the luminosity information of all the four sources providing an excess into IceCube data rather than assume all the sources to be NGC 1068-like. However, within all uncertainties the results are consistent. The CR cut-off at $200\, \rm TeV$ naturally suppresses the contribution of seyfert galaxies  at $E_{\nu} \ge 100\, \rm TeV$, leaving room for other astrophysical sources such blazars (see \cite{Padovani:2024tgx,Palladino:2018lov} for instance) or starburst galaxies \cite{Ambrosone:2020evo,Ambrosone:2024xzk,Peretti:2019vsj}. 
However, the chosen cut-off is only driven by the point-like IceCube observations which sets an upper limit of neutrino emission at $\sim 10-50\, \rm TeV$ rather than first-principle calculations. Therefore, future theoretical studies directly probing the high-energy CR cut-offs in these sources are fundamental. In fact, a higher $E_{\rm max}$ value would lead to  a higher diffuse neutrino flux at $E_{\nu}\sim 100\, \rm TeV$, providing further constraints to seyfert galaxies emissions.
We also highlight that the upcoming KM3NeT telescope~\cite{KM3NeT:2024uhg} will be fundamental in order to probe the role of each of these components into the high-energy neutrino sky especially in the $1-10\, \rm TeV$ energy range. 
Indeed, shower-like events, having a reduced background rate will be a perfect sample to investigate the diffuse emission of these galaxies with.

\section{Conclusions}\label{sec:conclusions}

In this paper, we have revisited the constraints on the neutrino emission of seyfert galaxies, initially proposed by V.S. Berezinsky~\cite{1981ICRC....1..238B}, exploiting the latest observations by the IceCube collaboration~\cite{Abbasi:2024hwv,Abbasi:2024ofy,IceCube:2023tts,IceCube:2023nai,Neronov:2023aks}. 
We develop a theoretical model accounting both for CR escape and energy losses mechanisms, assuming equi-ripartition between CR and magnetic energy densities. The four point-like sources providing an excess into IceCube data are consistent with an equi-ripartion hypothesis but the beta plasma parameter required to fit the data is rather low $(\beta \lesssim 1)$, pointing to a high energetics carried by CRs inside these sources. Therefore, future dedicated analyses will be crucial to investigate if such condition can be met in environment of seyfert galaxies.
Furthermore, extrapolating the neutrino emission to the whole source population might overestimate the diffuse neutrino flux at $\sim 1-10\, \rm TeV$ energies leading to a potential tension in the high-energy neutrino sky. We might argue that all the sources might not be in  an equi-ripartition state and if we take a back-of-the-envelope estimate of $\sim 1/10$ of the sources~(considering that IceCube has observed an excess for 3 sources out of a catalogue of the 30 most luminous sources  in the northern hemisphere), this would sensibly reduce the tension (see also Ref.~\cite{Padovani:2024tgx} for further remarks). All in all, all these results point to the fact that regions sourrounding the SBMHs in AGNs might sensibly produce high-energy neutrinos.

\acknowledgments

The authors are supported by the research project TAsP (Theoretical Astroparticle Physics) funded by the Istituto Nazionale di Fisica Nucleare (INFN).


\bibliography{references}


\bibliographystyle{unsrt}

\appendix

\section{Details on the Dynamical Timescales}\label{sec:timescales}
In this section, we report details on the dynamical timescale in the coronae.
The escape timescale is given by the in-fall timescale onto the black hole  $\tau_{\rm esc} = R/V_{\rm fall}$, with $V_{\rm fall} = \alpha \sqrt{2GM_{\rm BH}/R}$. Therefore $\tau_{\rm esc}$ reads
\begin{equation}
    \tau_{\rm esc} = 3.54\cdot 10^{6} \bigg(\frac{\alpha}{0.1}\bigg) \bigg(\frac{r}{20}\bigg)^{1/2} \bigg( \frac{M_{\rm BH}}{10^8\, \rm M_{\odot}}\bigg) \, [s]
\end{equation}
we fix the friction coefficient $\alpha = 0.1$ \cite{Inoue:2024nap}. 
For the mass of the SMBHs, we make use of \cite{Inoue:2019fil,Mayers:2018hau}
\begin{equation}
 M_{\rm BH} = 2\cdot 10^{7}\, M_{\odot} \bigg(\frac{L_X}{1.155\cdot 10^{43}\, \rm erg\, \rm s^{-1}}\bigg)^{0.746}
\end{equation}
in this way, we can express $M_{\rm BH}$ in terms of $L_X$. 
$\tau_{\rm loss}$ is given by the competition between inverse compton (IC), Synchrotron, Bethe-Heitler pair production, photomeson production and pp collisions. For IC and Synchrotron the timescales read \cite{Inoue:2019fil}

\begin{equation}
    \tau_{\rm IC,\rm syn}(E) = \frac{3}{4}\bigg(\frac{m_p}{m_e}\bigg)^3 \frac{m_e c^2}{c\sigma_T U_{\rm ph, B}} \bigg(\frac{E}{m_p c^2}\bigg)^{-1}
\end{equation}
where $U_{\rm ph} = L_{\rm ph}/(4\pi R^2 c)$ is the total background photon background with $L_{\rm ph}$ is the bolometric luminosity. $U_B = B^2/8\pi$ is the magnetic energy density.  The timescale for Bethe-Heitler pair production reads~\cite{Inoue:2019fil}

\begin{equation}
    \tau_{\rm BH}^{-1} = \frac{7 (m_ec^2)^3 \alpha_f \sigma_T \cdot c}{9\sqrt{2} m_p c^2} \bigg(\frac{E}{m_p c^2}\bigg)^{-2} \int_{m_e m_p c^4/E}^{+\infty} d\epsilon \frac{n_{\rm ph}(\epsilon)}{\epsilon^3} \cdot \bigg[ \bigg(\frac{2 E \cdot \epsilon}{m_e m_p c^4}\bigg)^{3/2}\bigg(\rm log \bigg(\frac{2 E \cdot \epsilon}{m_e m_p c^4} \bigg) -\frac{2}{3}\bigg) + \frac{2}{3}\bigg]
\end{equation}
where $n_{\rm ph}(\epsilon)$ is the background photon density from Ref.~\cite{Inoue:2019fil}. For photomeson interaction, we use \cite{2012JCAP...11..058G}

\begin{equation}
    \tau_{p\gamma}^{-1} = \frac{c^5 m_p^2}{2 E^2} \int_0^{+\infty} dE^{'} \frac{n_{\rm ph}(E^{'})}{E^{'3}} \int_{E_{\rm th}}^{2 \frac{E\cdot E^{'}}{m_p c^2}} d\epsilon \epsilon \sigma_{p\gamma}(\epsilon) K_{p\gamma}(\epsilon) 
\end{equation}
where $\sigma_{p\gamma}(\epsilon)$ is total inelastic $p\gamma$ cross section in the proton rest frame \cite{Kelner:2008ke}. $K_{p\gamma}$ is approximated with a step function \cite{2012JCAP...11..058G}
\begin{equation}
 K_{p\gamma} (\epsilon) = 
    \begin{cases*}
        0.2 & if $\epsilon < 1\, \rm GeV $\\
        0.6 & otherwise
    \end{cases*}
\end{equation}
finally, $E_{\rm th} = 145\, \rm MeV$ is the kinetic threshold for the process. The pp timescale reads \cite{Inoue:2019fil,Kelner:2006tc}
\begin{equation}
    \tau_{pp} = \frac{1}{k_p n_p \sigma_{pp} c}
\end{equation}
where $k_p = 0.5$ is the mean inelasticity of the process.  $n_p$ is the gas density and it is set considering charge neutrality of the corona as \cite{Inoue:2019fil}

\begin{equation}
    n_p = \frac{\tau_{\tau}}{R\sigma_T} \simeq  2.8 \cdot 10^9 \frac{\tau_{\tau}}{1.1}\bigg(\frac{r}{20}\bigg)^{-1} \bigg(\frac{M_{\rm BH}}{10^8\, \rm M_{\odot}}\bigg)^{-1}\, [\rm cm^{-3}]
\end{equation}
where $\tau_{\tau} = 1.1$ is the Thompson scattering optical depth \cite{Inoue:2019fil}.  We can analytically estimate the efficiency to produce neutrinos as
$F_{\rm cal} \simeq (\tau_{\rm loss}^{-1} + \tau_{esc}^{-1})^{-1}/\tau_{pp}$, $F_{\rm cal} \simeq (\tau_{\rm loss}^{-1} + \tau_{esc}^{-1})^{-1}/\tau_{p\gamma}$ respectively for $pp$ and $p\gamma$ interactions \cite{Inoue:2019fil}. 

\begin{figure}[h!]
    \centering
    \includegraphics[width=0.49\columnwidth]{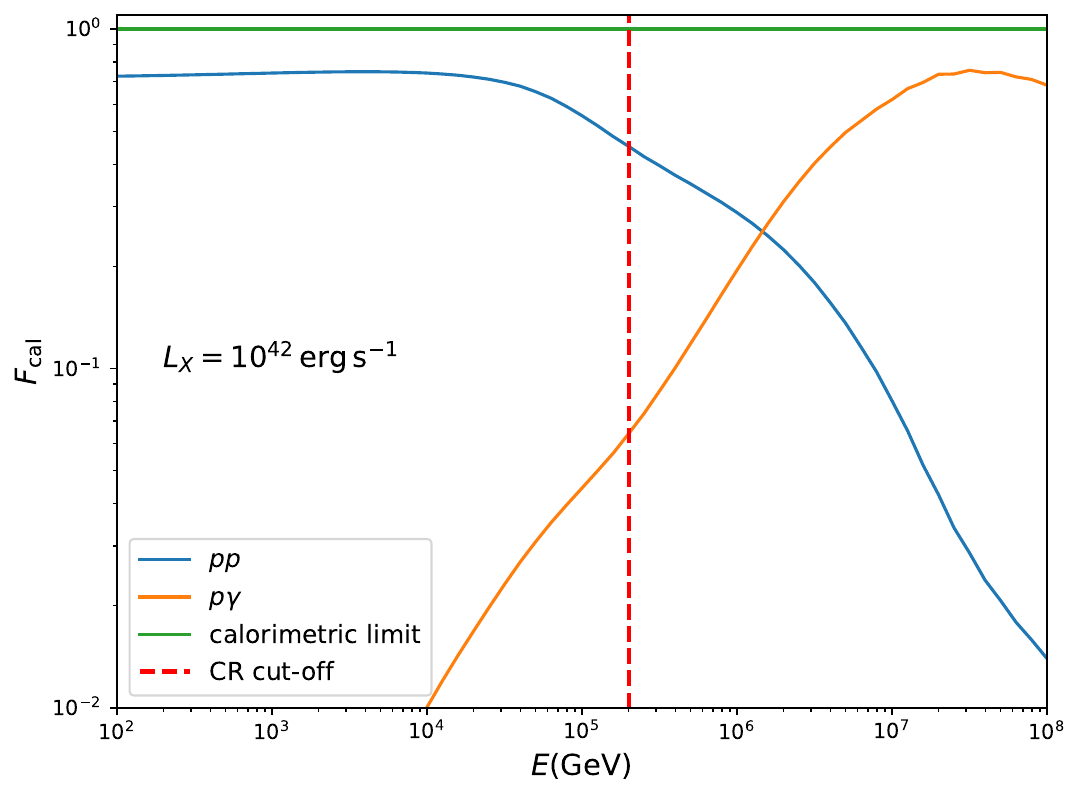}
     \includegraphics[width=0.49\columnwidth]{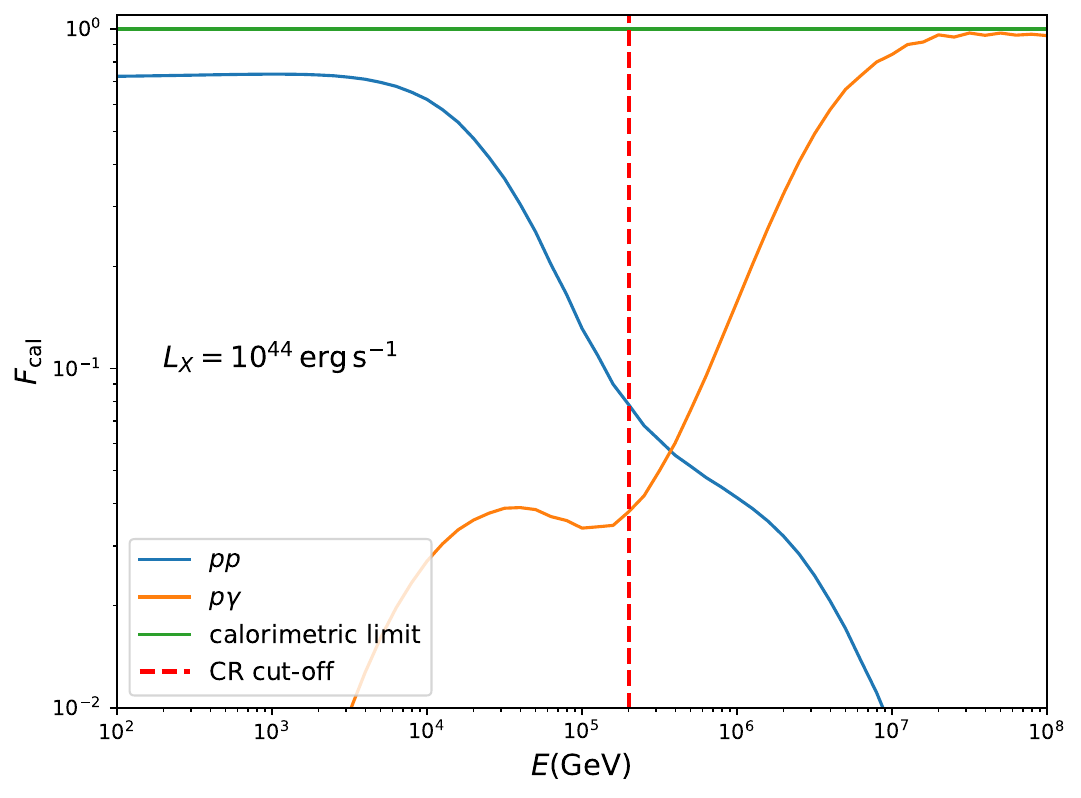}
     \includegraphics[width=0.49\columnwidth]{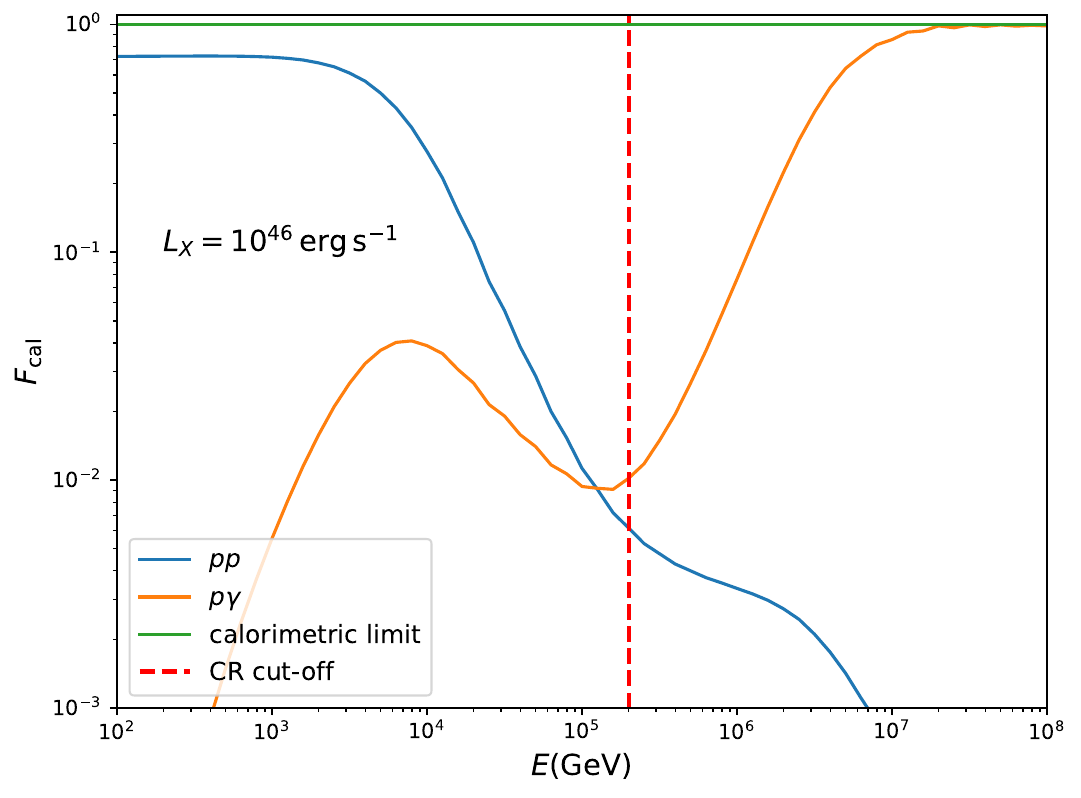}
    \caption{efficiency for neutrino production as a function of the CR energy. The blue line corresponds to the $pp$ collision channel, the orange line corresponds to $p\gamma$ interactions and the green line corresponds to the calorimetric limit. We also report a dashed red vertical line corresponding to the CR cut-off assumed in the analysis. The top left, top right and the bottom panels respectively  corresponds to $L_X = 10^{42}\, \rm erg\, \rm s^{-1}$, $L_X = 10^{44}\, \rm erg\, \rm s^{-1}$ and $L_X = 10^{46}\, \rm erg\, \rm s^{-1}$.}
    \label{fig:efficiency}
\end{figure}

Fig. \ref{fig:efficiency} shows the efficiency for three different values of $L_X$, $10^{42}\, \rm erg\, \rm s^{-1}$, $10^{44}\, \rm erg\, \rm s^{-1}$, $10^{46}\, \rm erg\, \rm s^{-1}$ as a function of the CR energy, fixing $\beta = 0.32$ as the best-fit scenario obtained in the main text.

The results show that for CR energy below the cut-off, the neutrino production is dominated by pp collisions.



\end{document}